\documentclass[english,twocolumn,prl,superscriptaddress]{revtex4-1}
\usepackage{graphicx}
\usepackage{bm}
\usepackage{braket}
\usepackage{amsmath}
\usepackage{amsthm}
\usepackage{commath}
\usepackage{dsfont}
\usepackage{xcolor}
\usepackage{extarrows}
\usepackage{nicefrac}
\usepackage{siunitx}
\usepackage{braket}
\usepackage{todonotes}

\newcommand{\circled}[1]{\raisebox{.5pt}{\textcircled{\raisebox{-.9pt} {#1}}}}

\begin{document}
\title{Topological Spin Texture of Chiral Edge States
in Photonic \\
Two-Dimensional Quantum Walks}
\author{Chao Chen}
\altaffiliation{These authors contributed equally to this work.}
\affiliation{Hefei National Laboratory for Physical Sciences at the Microscale and Department of Modern Physics, University of Science and Technology of China, Shanghai Branch, Shanghai 201315, China}
\affiliation{CAS-Alibaba Quantum Computing Laboratory, CAS Centre for Excellence in Quantum Information and Quantum Physics, Shanghai 201315, China}
\affiliation{National Laboratory of Solid State Microstructures, School of Physics, Nanjing University, Nanjing
210093, China}

\author{Xing Ding}
\altaffiliation{These authors contributed equally to this work.}
\affiliation{Hefei National Laboratory for Physical Sciences at the Microscale and Department of Modern Physics,
University of Science and Technology of China, Shanghai Branch, Shanghai 201315, China}
\affiliation{CAS-Alibaba Quantum Computing Laboratory, CAS Centre for Excellence in Quantum Information and Quantum Physics, Shanghai 201315, China}

\author{Jian Qin}
\altaffiliation{These authors contributed equally to this work.}
\affiliation{Hefei National Laboratory for Physical Sciences at the Microscale and Department of Modern Physics, University of Science and Technology of China, Shanghai Branch, Shanghai 201315, China}
\affiliation{CAS-Alibaba Quantum Computing Laboratory, CAS Centre for Excellence in Quantum Information and Quantum Physics, Shanghai 201315, China}

\author{Jizhou Wu}
\altaffiliation{These authors contributed equally to this work.}
\affiliation{Hefei National Laboratory for Physical Sciences at the Microscale and Department of Modern Physics,
University of Science and Technology of China, Shanghai Branch, Shanghai 201315, China}
\affiliation{CAS-Alibaba Quantum Computing Laboratory, CAS Centre for Excellence in Quantum Information and Quantum Physics, Shanghai 201315, China}

\author{Yu He}
\affiliation{Shenzhen Institute for Quantum Science and Engineering, Southern University of Science and Technology, Shenzhen 518055, China}

\author{\\Chao-Yang Lu}
\affiliation{Hefei National Laboratory for Physical Sciences at the Microscale and Department of Modern Physics,
University of Science and Technology of China, Shanghai Branch, Shanghai 201315, China}
\affiliation{CAS-Alibaba Quantum Computing Laboratory, CAS Centre for Excellence in Quantum Information and Quantum Physics, Shanghai 201315, China}

\author{Li Li}
\email{eidos@ustc.edu.cn}
\affiliation{Hefei National Laboratory for Physical Sciences at the Microscale and Department of Modern Physics,
University of Science and Technology of China, Shanghai Branch, Shanghai 201315, China}
\affiliation{CAS-Alibaba Quantum Computing Laboratory, CAS Centre for Excellence in Quantum Information and Quantum Physics, Shanghai 201315, China}

\author{Xiong-Jun Liu}
\email{xiongjunliu@pku.edu.cn}
\affiliation{International Center for Quantum Materials and School of Physics, Peking University, Beijing 100871, China}
\affiliation{CAS Center for Excellence in Topological Quantum Computation, University of Chinese Academy of Sciences, Beijing 100190, China}
\affiliation{Shenzhen Institute for Quantum Science and Engineering, Southern University of Science and Technology, Shenzhen 518055, China}

\author{Barry C. Sanders}
\email{sandersb@ucalgary.ca}
\affiliation{Hefei National Laboratory for Physical Sciences at the Microscale and Department of Modern Physics,
University of Science and Technology of China, Shanghai Branch, Shanghai 201315, China}
\affiliation{CAS-Alibaba Quantum Computing Laboratory, CAS Centre for Excellence in Quantum Information and Quantum Physics, Shanghai 201315, China}
\affiliation{Institute for Quantum Science and Technology, University of Calgary, Alberta T2N 1N4, Canada}

\author{Jian-Wei Pan}
\affiliation{Hefei National Laboratory for Physical Sciences at the Microscale and Department of Modern Physics,
University of Science and Technology of China, Shanghai Branch, Shanghai 201315, China}
\affiliation{CAS-Alibaba Quantum Computing Laboratory, CAS Centre for Excellence in Quantum Information and Quantum Physics, Shanghai 201315, China}
\date{\today}

\begin{abstract}
Topological insulators host topology-linked boundary states, whose spin and charge degrees of
freedom could be exploited to design topological devices with enhanced functionality.
We experimentally observe that dissipationless chiral edge states in a spin-orbit coupled anomalous Floquet topological phase exhibit topological spin texture on boundaries,
realized via a two-dimensional quantum walk.
Our experiment shows that,
for a walker traveling around a closed loop
along the boundary in real space,
its spin evolves and winds through a great circle on the Bloch
sphere,
which implies that edge-spin texture has nontrivial winding.
This topological spin winding is protected by a chiral-like symmetry emerging for the low-energy Hamiltonian.
Our experiment confirms that
two-dimensional anomalous Floquet topological systems exhibit topological spin texture on the boundary,
which could inspire novel topology-based spintronic phenomena and devices.
\end{abstract}
\maketitle

Topological materials host protected boundary states that could have applications to topology-based
electronic devices~\cite{hasanColloquiumTopologicalInsulator2010,qiTopologicalInsulatorsSuperconductors2011,fanSpintronicsBasedTopological2016,pesinSpintronicsPseudospintronicsGraphene2012,smejkalRouteDiracWeyl2017,smejkalTopologicalAntiferromagneticSpintronics2018,rojas-sanchezSpinChargeConversion2016}.
For example, time-reversal
invariant topological insulators (TIs) have helical boundary modes that are spin-momentum locked due to
strong spin-orbit coupling and can be exploited to
generate spin currents~\cite{murakamiDissipationlessQuantumSpin2003,sinovaUniversalIntrinsicSpin2004,sinovaSpinHallEffects2015} by applying an external field.
Fabrication of such TI surfaces and magnetic materials could facilitate control of magnetic domains via
spin-transfer torque induced through coupling to
surface spin currents~\cite{mellnikSpintransferTorqueGenerated2014,wangTopologicalSurfaceStates2015},
thereby realizing a key building
block of spin-based logic devices~\cite{zuticSpintronicsFundamentalsApplications2004,wolfSpintronicsSpinBasedElectronics2001}.
Breaking time-reversal
symmetry with ferromagnetic ordering could drive the
TI into a quantum anomalous Hall~(QAH) insulator~\cite{haldaneModelQuantumHall1988} in
the two-dimensional (2D) regime, in which case TI
helical edge states transform into chiral modes~\cite{liuQuantumAnomalousHall2016}.
QAH insulators are widely studied in solid-state systems~\cite{qiGeneralTheoremRelating2006a,liuQuantumAnomalousHall2008,yuQuantizedAnomalousHall2010,changExperimentalObservationQuantum2013,bestwickPreciseQuantizationAnomalous2015,dengQuantumAnomalousHall2020,dengHightemperaturequantumanomalous2021a,sunRationalDesignPrinciples2019}
and also ultracold atoms~\cite{wuOrbitalAnalogueQuantum2008,liuQuantumAnomalousHall2010,liuRealization2DSpinOrbit2014,zhengFloquetTopologicalStates2014,jotzuExperimentalRealizationTopological2014,aidelsburgerMeasuringChernNumber2015,wuRealizationTwodimensionalSpinorbit2016,goldmanTopologicalQuantumMatter2016,wangDiracRashbaWeyltype2018,sunUncoverTopologyQuantum2018,sunHighlyControllableRobust2018}.
For a two-band spin-orbit-coupled QAH insulator,
chiral edge states are spin polarized~\cite{zhangElectricallyTunableSpin2016,perez-piskunowHingeSpinPolarization2021,tanTwoParticleBerryPhase2022}. 
Furthermore, when the low-energy Hamiltonian has chiral-like symmetry, the corresponding chiral edge mode exhibits exotic topological spin texture over the one-dimensional (1D)
closed boundary~\cite{wuTopologicalSpinTexture2014}, which is
observed experimentally here for the first time.

Here, we investigate spin texture of chiral edge states in a 2D Floquet quantum walk system. 
Though the Chern number of the effective Hamiltonian vanishes,
our 2D discrete-time quantum walk (2DQW) model is topologically nontrivial when a winding number $W$ defined in time-and-momentum space is nonzero~\cite{rudnerAnomalousEdgeStates2013,grohRobustnessTopologicallyProtected2016,maczewskyObservationPhotonicAnomalous2017a,mukherjeeExperimentalObservationAnomalous2017,zhangUnifiedTheoryCharacterize2020,winterspergerRealizationAnomalousFloquet2020}. 
Building on previous experiments~\cite{kitagawaExploringTopologicalPhases2010,kitagawaTopologicalPhenomenaQuantum2012,schreiber2DQuantumWalk2012,crespiAndersonLocalizationEntangled2013,flurinObservingTopologicalInvariants2017,chenObservationTopologicallyProtected2018,wangExperimentalObservationTopologically2018,barkhofenSupersymmetricPolarizationAnomaly2018,wangSimulatingDynamicQuantum2019,weidemannTopologicalFunnelingLight2020},
we push the modulation speed and evolution time of our setup to the limit to create an inhomogeneous square
lattice, 
which can be divided into an outer region and an inner region of $9\times13$ sites. Chiral edge modes exist
on the 1D closed boundary when the topological phases of the inner
and outer regions are distinct. In particular, we introduce spin tomography to map out the real-time spin vector of the walker when it moves along the 1D closed boundary.

We successfully observe that, when a walker travels over the whole closed
boundary, its spin evolves and realizes a spin winding in the $\sigma_x$-$\sigma_z$ plane of the Bloch
sphere, 
which
is linked to chiral-like symmetry of our low-energy Hamiltonian.
Opposite spin winding of chiral edge states controlled by tuning experimental parameters is confirmed by our experiment. 
The local spins on the edge show small fluctuation due to experimental imperfections; however, the overall spin winding is robust.
Our
experiment yields insight into manipulating the
spin degree of freedom through dissipationless topological edge states.
Topological spin texture of chiral edge states can be
applied to generate spin-polarized currents,
which could have applications for versatile spintronic devices such
as efficient spin generators and spin filters~\cite{wuTopologicalSpinTexture2014}.

We commence with describing the model for our experiment.
2DQWs describe the evolution of a spin-$\nicefrac12$ particle on a square lattice under repeated applications of a unitary step operation comprising spin rotation and spin-dependent translation.
Given the basis vector $\ket{x,y,s}$
for internal spin state~$s$ and walker coordinates $x,y$ on the 2D lattice,
the unitary step operator is
\begin{align}
U(\bm{\theta})=T_y R(\theta_2) T_x R(\theta_1)\label{eq:1step}
\end{align}
for $\bm{\theta}=(\theta_1,\theta_2)$ the parameter space and
\begin{align}
 R(\theta_i)=e^{-\text{i}\frac{\theta_i}{2}\sigma_y},\,
 \sigma_y
 =\begin{pmatrix}0&-\text{i}\\\text{i}&0\end{pmatrix},
\end{align}
a spin rotation around the~$y$ axis.
Spin-dependent translation in position space is
\begin{eqnarray}
 T_x=\sum_x \left[\ket{x+1}\bra{x}\otimes \ket\uparrow\bra\uparrow
 +\ket{x-1}\bra{x}\otimes \ket\downarrow\bra\downarrow\right]
\end{eqnarray}
and $T_y=T_x(x\to y)$ denotes translation in the~$y$ direction.

\begin{figure}
\includegraphics[width=\columnwidth]{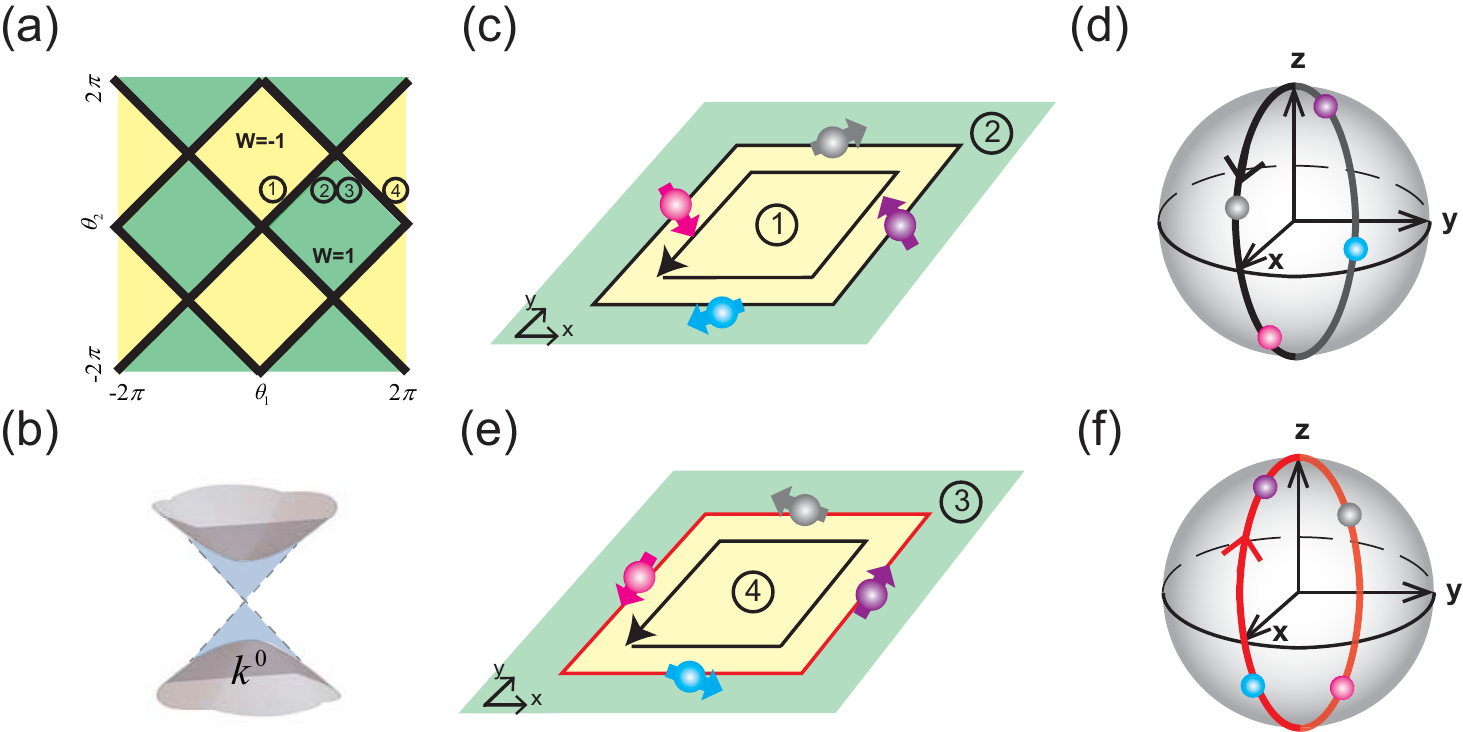}
\caption{Phase diagram, band structure, and sketches
of spin texture for chiral edge states.
(a)~Phase diagram of 2DQWs in parameter space~$\bm{\theta}$.
Two distinct topological phases are present, whose winding numbers are~$-1$ (yellow) and~$1$ (green), respectively.
Black lines denote phase boundaries. (b)~Schematic band structure.
When choosing $\bm{\theta}$ at phase boundaries,
the band gap closes at $\bm{k}^0$ in momentum space.
(c)~spin texture on a rectangular edge when the inner and outer regions are ruled by $\theta_1^\text{in}=\nicefrac{\pi}6,\theta_1^\text{out}=\nicefrac{5\pi}6$, which are represented by \circled1 and \circled2, respectively.
$\theta_2^\text{in/out}$ are always fixed at $\nicefrac{\pi}2$.
Chirality of the edge states is indicated by the black arrowed line along the edge. (d)~Spin winding on the Bloch sphere of spin texture shown in~(c).
(e)~Schematic spin texture on a rectangular edge when the inner region and outer region are ruled by $\theta_1^\text{in}=\nicefrac{11\pi}6,\theta_1^\text{out}=\nicefrac{7\pi}6$, represented by \circled4 and \circled3, respectively. (f)~Spin winding on the Bloch sphere of spin texture shown in~(e).}
\label{fig:1}
\end{figure}

In terms of Floquet band theory for $U(\theta_1,\theta_2)$,
the effective Hamiltonian in momentum space~\cite{chenObservationTopologicallyProtected2018} is
\begin{equation}
 H_\text{eff}(\bm{\theta})
 =\int d^2\bm{k}
 \underbrace{\left[E\left(\bm{k}\right)\bm{n}\left(\bm{k}\right)\cdot\bm\sigma \right]}_{H_\text{eff}(\bm{k})}\otimes\ket{\bm{k}}\bra{\bm{k}},
\label{eq:hamiltonian}
\end{equation}
with
$\bm\sigma=(\sigma_x,\sigma_y,\sigma_z)$ the Pauli matrix vector
and $\bm\theta$ present but notationally suppressed in~$E(\bm{k})$ and~$\bm{n}(\bm{k})$.
Here, we explore the chiral edge states at the interface between two different phases in the phase diagram~[Fig.~\ref{fig:1}(a)] with topological invariant $W=+/-1$, respectively, as tuned by manipulating the parameters $\bm{\theta}$ .

To investigate the properties of edge states that emerge at low energy, we make a low-energy approximation of the bulk Hamiltonian around~$\bm{k}^0$,
which are the gap closing points in momentum space when~$\bm{\theta}$ is set at topological phase-transition boundaries as shown in Fig.~\ref{fig:1}(b).
In the 2DQW model, when edge states correspond to the phase transition across the $45^{\circ}$-diagonal phase boundaries, where the gap closing points are at~$\bm{k}^0=(\pm\nicefrac{\pi}2,\mp\nicefrac{\pi}2)$, we obtain the approximate low-energy Hamiltonian around $\bm{k}^0$ as
$\tilde{H}(\bm{n})
=\bm{n}\cdot\bm{\sigma}/\text{sinc}(\theta_-)$ for $\theta_\pm:=\left(\theta_1\pm\theta_2\right)/2$ and
\begin{equation}
\label{eq:nxnynz}
\bm{n}
=\begin{pmatrix}
\sin{\theta_+}\delta k_x +\sin{\theta_-}\delta k_y\\
\sin{\theta_-}\frac{2-\delta k_x^2-\delta k_y^2}{2}-\sin{\theta_+}\delta k_x\delta k_y\\
-\cos{\theta_+}\delta k_x -\cos{\theta_-}\delta k_y
\end{pmatrix}
\end{equation}
with $\abs{\bm{\delta k}}$ the deviation from $\bm{k}^0$.
Contrariwise, when corresponding to $-45^{\circ}$-diagonal phase boundaries,
band-gap closing points are
$\bm{k}^0\in\{(\pm\pi,0),(0,\pm\pi)\}$,
which exchanges $\theta_+\leftrightarrow \theta_-$
in the approximate Hamiltonian
(see Supplemental Material~\cite{supplemental_material}  for details).

Now we analyze symmetry of the low-energy Hamiltonian for chiral edge states.
In our experiments, we set $\theta_{2}=\nicefrac{\pi}{2}$. 
In this case, our low-energy Hamiltonian satisfies chiral-like symmetry ($\mathcal{S}\tilde{H}=-\tilde{H}\mathcal{S}$), the central ingredient for the emergence of topological spin texture~\cite{wuTopologicalSpinTexture2014,supplemental_material}. 
Consider a topological 2DQW that is periodic along the $x$ direction and has open boundary in the $y$ direction.
We have chiral-like symmetry operator $\mathcal{S}=C_{x}\mathcal{M}_{x}$ for low-energy Hamiltonian in Eq.~(\ref{eq:nxnynz}), with $C_x=\cos{\theta_-}\sigma_x+\sin{\theta_-}\sigma_z$ and $\mathcal{M}_{x}$ the spatial reflection operator along the $x$ direction. Having the chiral-like symmetry, the edge states with low energies on two opposite edges 
are the mutually orthogonal eigenstates of the $C_x$, akin to the end states of a 1D chiral TI~\cite{wuTopologicalSpinTexture2014}. 
Moreover, when the direction of the edge changes, the spin on the edge changes with the edge in the same way, which gives rise to the topological spin winding on a closed edge~\cite{supplemental_material}.

The chiral-like symmetry does not involve the $\sigma_y$ component; thus, low-energy edge states are always polarized in the $\sigma_x$-$\sigma_z$ plane. Define spin winding~$w^{s}$ as the winding of a walker's spin on the Bloch sphere when the walker encircles the closed edge.
$w^s$ is $\pm 1$ when the edge-spin texture corresponding to $\pm45^{\circ}$ phase boundaries, respectively~(see Supplemental Material~\cite{supplemental_material}). 
Notice that here $\mathcal{S}$ is $\theta_{1}$-dependent; thus, the spin vector along the edge also depends on $\theta_{1}$, which is different from the earlier case~\cite{wuTopologicalSpinTexture2014}.
However, the topological spin winding number
is independent of $\theta_1$ and protected by chiral-like symmetry, hence robust~\cite{supplemental_material}.
For edge states of high energy, low-energy Hamiltonian is not valid, and spin polarization does not strictly lie in $\sigma_x$-$\sigma_z$ plane\cite{supplemental_material}.

Experimentally, we measure spin texture on the rectangular edge shown in Figs.~\ref{fig:1}(c) and (e).
In both settings,
the topological invariants of the inner and outer bulks are set to be $W=-1$ and $W=1$, respectively.
Because the anticlockwise-moving edge states at the interface between phases \circled1 and \circled 2 (\circled 3 and \circled 4) correspond to $45^{\circ}$ ($-45^{\circ}$) phase boundaries,
the two settings have opposite $w^{s}$ as shown in Figs.~\ref{fig:1}(d) and~\ref{fig:1}(f).
Governed by different low-energy Hamiltonians, spin polarization on the outer and inner sides of the edge is different and can be measured independently. Here, we only focus on the spin texture of the outer edge, but similar analysis pertains for spin on the inner edge~\cite{supplemental_material}. For the setting in Fig.~\ref{fig:1}(c),
chiral-like symmetry with $C_x=\cos{\frac{\pi}6}\sigma_x+\sin{\frac{\pi}6}\sigma_z$ ($C_y=\cos{\frac{2\pi}{3}}\sigma_x+\sin{\frac{2\pi}{3}}\sigma_z$) holds for low-energy Hamiltonian with edges along $x$ ($y$) direction. Similarly, for the setting in Fig.~\ref{fig:1}(e),
we have $C_x=\cos{\frac{5\pi}6}\sigma_x+\sin{\frac{5\pi}6}\sigma_z$ and $C_y=\cos{\frac{\pi}{3}}\sigma_x+\sin{\frac{\pi}{3}}\sigma_z$,
where the sign of the coefficient before term $\sigma_x$ gets flipped~\cite{supplemental_material}.

\begin{figure}
 \includegraphics[width=\columnwidth]{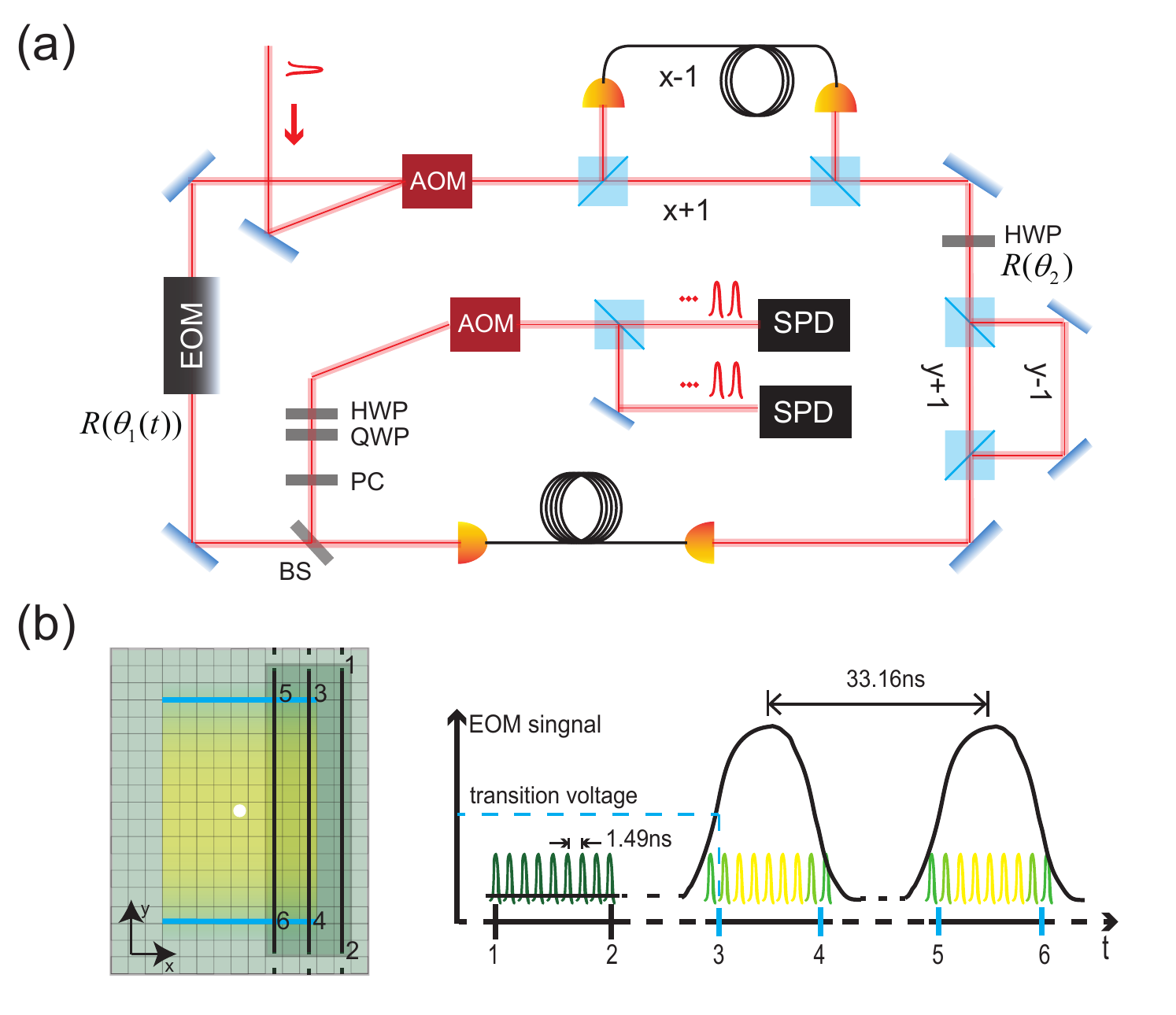}
 \caption{Experimental realization of 2DQWs with a rectangular edge.
 (a)~Experimental setup.
 An $\SI{895}{\nm}$ laser pulse with $\SI{30}{\ps}$ pulse width
 is coupled into the main optical loop of by an AOM.
 This pulse evolves into a pulse sequence after going through a $\SI{6}{\m}$ PD fiber delay, a HWP, a $\SI{45}{\cm}$ free-space optical delay,
 and a fast-shifting EOM successively and repeatedly.
 The pulse sequence is coupled out by a beam splitter, projected in the $\sigma_{x,y,z}$ bases
 and measured by single-photon detectors (SPDs).
 PC, phase compensator.
 (b)~Mapping between the position space and the temporal space.
 The walker's evolution in green (outer) and yellow (inner) regions belongs to distinct topological phases.
 The coordinate of the white point is (0,0).
 Guided by markers 1--6,
 each time bin finds its correspondence in position space and is also colored in green or yellow accordingly.
 A modulated voltage is applied on the EOM to implement effective $R(\theta_1^\text{in (out)})$ for yellow (green) time bins.
 Blue lines denote two edges parallel to the~$x$ axis.
The rising and falling edge of EOM signal is about \SI{5}{\ns}.}
 \label{fig:2}
\end{figure}

In our experiment, we measure spin polarization of a walker traveling over a $9\times13$ sized closed rectangular edge in real space. Here, we devise a laser pulse coherently distributed
into different temporal modes after repeatedly going through a nested fiber-loop setup [Fig.~\ref{fig:2}(a)]
to simulate a quantum walk on a periodic square lattice.
The walker's spin degree of freedom
is manifested by the laser-pulse optical polarization
with spin up (down) represented by horizontal (vertical) polarization.
Elements of one walking step, Eq.~\eqref{eq:1step}, namely
$R(\theta_1)$,
$T_x$,
$R(\theta_2)$, and~$T_y$,
are implemented successively
by a fast-switching electro-optic modulator~(EOM),
a polarization-dependent (PD) optical delay,
a half-wave plate~(HWP),
and another PD optical delay.
The operation~$R(\theta_1)$ of the first step is incorporated into the initial polarization preparation.
All optical elements are finely tuned, and decoherence is negligible due to the stability of the optical delays 
during the evolution time in the circuit of each incident laser pulse (about \SI{30}{\us}).

After loops of traveling, laser pulses split off from the initial laser pulse are attenuated to the single-photon level
and are coupled out passively by a 3\%-reflectivity beam splitter for measurement.
The measurement consists of reading out spin polarization and tagging the arrival time of each pulse.
Each photon's polarization is measured in $\sigma_{x,y,z}$ basis by a set comprising a HWP, a quarter-wave plate~(QWP), and a polarization beam splitter~(PBS).
Arrival time is recorded by single-photon detectors of time resolution $\SI{50}{\ps}$ with the assistance of an acoustic-optical modulator~(AOM) serving as an optical switch to remove undesired pulses.

To effectively implement inhomogeneous 2D quantum walk with a rectangular edge [Fig.~\ref{fig:2}(b)], the mapping between temporal axis and spatial coordinate must be established. The nearest two
time bins separated by \SI{1.49}{\ns} correspond to two sites in the same column with a difference of 2 in the $y$ axis; similarly, two sequences of time bins separated by \SI{33.16}{\ns} represent two columns with a distance of 2 lattice sites in the $x$ axis. Compared with the previous experiment~\cite{chenObservationTopologicallyProtected2018}, apart from edges along the $y$ axis, we also need to create edges along the $x$ direction (blue lines), for which the complicated modulation with fast voltage switching between adjacent time bins of different $y$ coordinates is desired. Given the rising and falling time of our EOM, 
the varying of the voltage takes about 5ns, which is about three times the separation between two time bins.
Therefore, along the $x$ direction a blunt edge across several sites can be constructed. Taking the shadowed region in the left panel of Fig.~\ref{fig:2}(b) as an example, depending on which region the time bins belong to, voltages either related to $R(\theta_1^\text{in})$ or $R(\theta_1^\text{out})$ are applied on the EOM (guided by markers 1--6). Time-dependent EOM signals need to be modulated and precisely aligned.

\begin{figure}
 \includegraphics[width=\columnwidth]{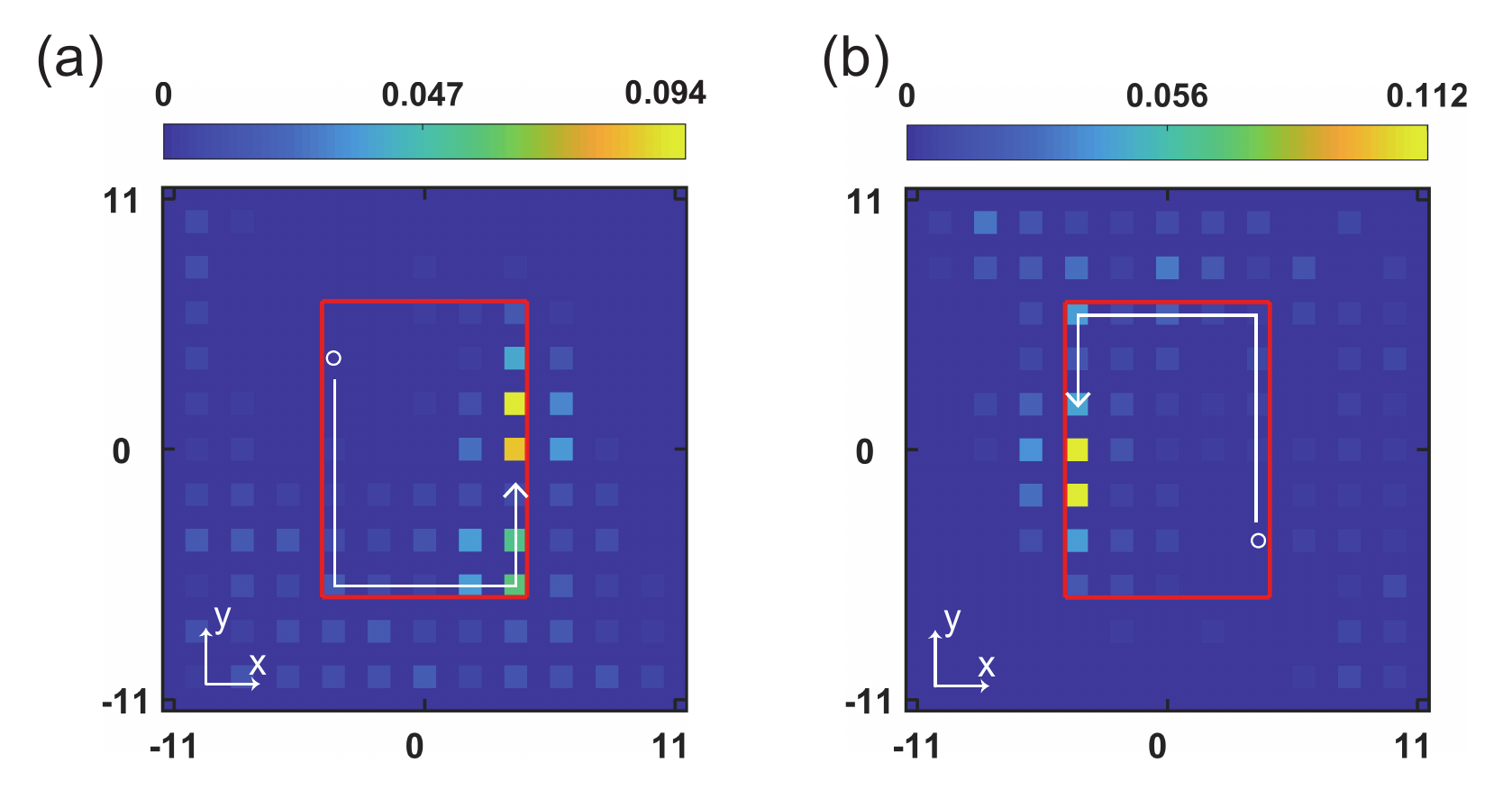}
 \caption{%
 Measured 2DQW probability distributions after 28 steps when $\theta_1^\text{in}=\nicefrac{\pi}6$,
 $\theta_1^\text{out}=\nicefrac{5\pi}6$, and $\theta_2^\text{in/out}=\nicefrac{\pi}2$. (a), (b) are the results of a walker starting at ($-4$,4) and (4,$-4$), marked by white circles, with spin initially polarized in~$V$ and~$H$, respectively.
 A $9\times13$ sized red rectangle centered at the origin shows the ideal boundary.
 White-arrow lines denote the evolution direction of the wave packet.
 Joining the paths in (a) and (b), the walker go around the whole boundary. }
 \label{fig:3}
\end{figure}

Here, we present experimental results for parameters with $\theta_1^\text{in}=\nicefrac{\pi}6$, $\theta_1^\text{out}=\nicefrac{5\pi}6$, and $\theta_2^\text{in/out}=\nicefrac{\pi}2$. The walker is initialized on the edge, and its spin is tuned to ensure the initial state has prominent overlap with edge states with $E\approx0$, while the wave packet in general is a superposition of many edge states of different quasimomenta.
Because the $\sigma_y$ values of the spin for the high-energy edge states with $\pm \Delta k$ momenta deviation from $\bm{k}^0$ point are opposite,
out-of-plane spin polarization of the superposed high-energy edge states cancels on average~\cite{supplemental_material}. 
The walker's wave function diffuses into the 2D lattice after several walking steps. However, with a probability about 0.3,
the walker is found to keep moving along the edge anticlockwise, which can be regarded as edge states.  The walker's probability distribution on the lattice after 28 walking steps is shown in Fig.~\ref{fig:3}. Combining the evolution path in Figs.~\ref{fig:3}(a) and (b),
the walker encircle the whole $9\times13$ rectangular edge. To describe the similarity between the measured probability distribution and theoretically simulated distribution, similarity is defined as $S=\sum_{x,y} \sqrt{P_{\text{th}}(x,y) P_{\text{exp}}(x,y)}$. For the measured $P_{\text{exp}}(x,y)$ shown in Figs.~\ref{fig:3}(a) and (b), their similarities are $0.943\pm0.003$ and $0.938\pm0.003$, respectively.

\begin{figure}
\includegraphics[width=\columnwidth]{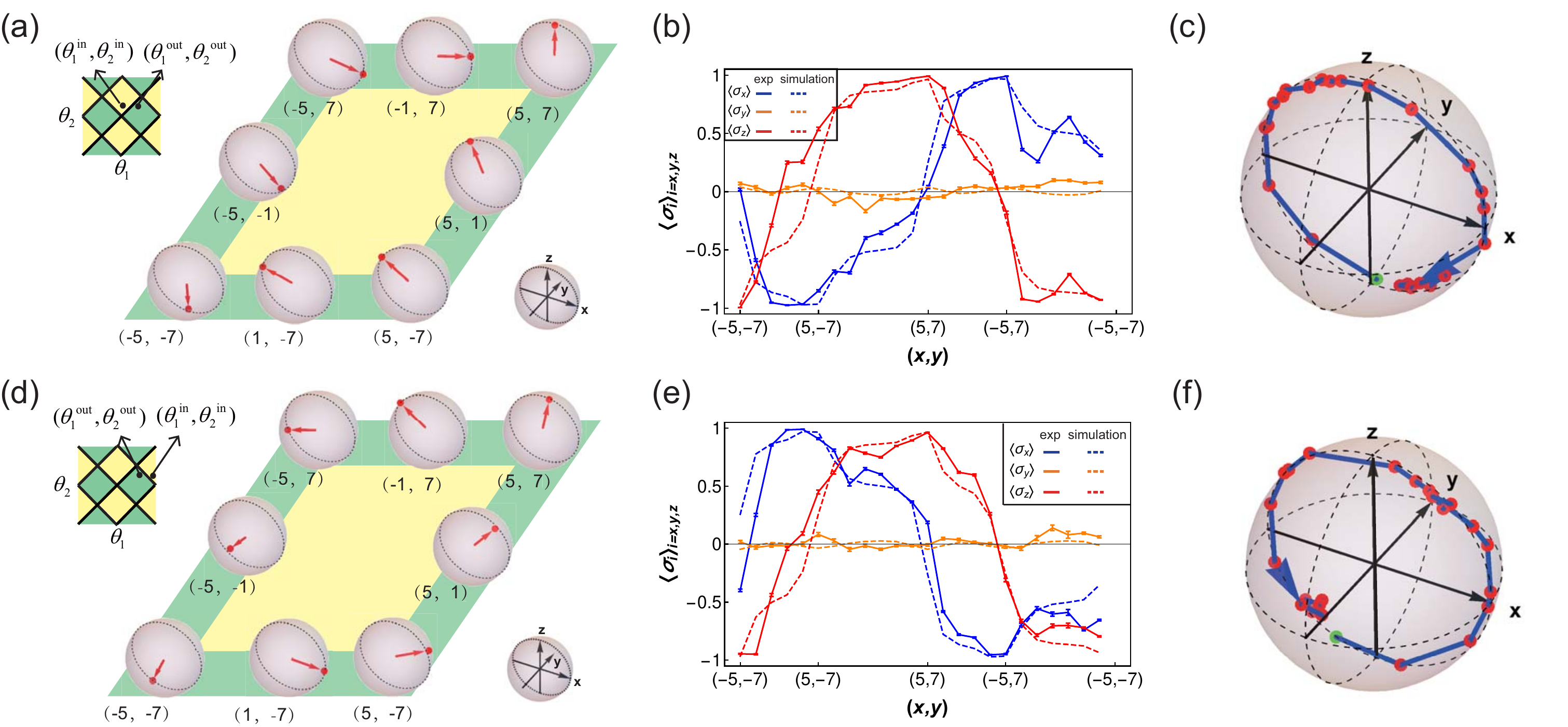}
\caption{%
Measured spin texture $\Braket{\bm{\sigma}}$ of spin on the edge and spin evolution on the Bloch sphere. (a)~Measured spin texture on the edge when $\theta_1^{\text{in}}=\nicefrac{\pi}6$ and $\theta_1^{\text{out}}=\nicefrac{5\pi}6$. The 2D lattice is divided into inner (yellow) and outer (green) regions, with $W^\text{in}=1$ and $W^\text{out}=-1$, respectively.
Measured spin polarization of the walker going through some typical positions on the edge are depicted as points on the Bloch sphere and directed by red vectors from the origin. The coordinates of the selected positions are labeled beside. (b)~Measured $\Braket{\sigma_x}, \Braket{\sigma_y}, \Braket{\sigma_z}$ of the walker's spin vs the walker's positions on the expanded edge. Numerically simulated results are plotted with dashed lines, and points with error bars denote experimentally measured values.
(c)~Spin evolution along the edge shown on the Bloch sphere. Reconstructed spin by pairs of ($\Braket{\sigma_x}, \Braket{\sigma_y}, \Braket{\sigma_z}$) in (b) are plotted as red points, except for the position ($-5$,$-7$), whose spin is colored in green. All spin points are connected by a blue arrowed line to indicate the spin-evolution direction. (d)--(f)~Experimental results as those in (a)--(c) for 2DQWs when $\theta_2^\text{in/out}=\nicefrac{\pi}2$ and $\theta_1^\text{in}=\nicefrac{11\pi}6$, $\theta_1^\text{out}=\nicefrac{7\pi}6$.}
\label{fig:4}
\end{figure}

As the walker moves along the boundary, its spin polarization evolves. As depicted in Fig.~\ref{fig:4}(a), antisymmetric spin texture is observed experimentally,
namely that spin is approximately polarized in the opposite direction for the walker going through two positions $(x,y)$ and $(-x,-y)$ on the edge. For clarity, we expand the rectangular edge into a 1D chain that starts from the lower left corner $(-5,-7)$,
proceeds to the lower right corner $(5,-7)$,
then follows the edge, and finally returns back to coordinate $(-5,-7)$,
shown as the horizontal axis of Fig.~\ref{fig:4}(b).
From the expanded edge, we sample 24 positions, with two neighboring ones having a spacing of 2 (See Supplemental Material~\cite{supplemental_material} for details).
Measured spin polarizations in the $\sigma_{x,y,z}$ basis vs those sampled positions are plotted in Fig.~\ref{fig:4}(b).
The numerical simulation is obtained by analyzing the edge state of the effective 2DQW Hamiltonian
on a $23\times23$ 2D lattice with a $9\times13$-sized topologically distinct region inside.
Experimentally measured values match well with simulated ones overall, where $\Braket{\sigma_y}$ vanishes along the whole edge. The main sources of imperfections in our experiments arise from deviations in calibrating optical delay and aligning the EOM signal.

Using state tomography, we reconstruct the spin state from each set of means~$\Braket{\bm{\sigma}}$
obtained by measurements.
In Fig.~\ref{fig:4}(c), the reconstructed spin states obtained from measured expectation values shown in Fig.~\ref{fig:4}(b) are represented by points almost on the Bloch sphere.
Those points are connected by a blue line in the order that their spatial coordinates are arranged on the expanded 1D chain.
When a walker goes through a closed path along the edge anticlockwise,
its spin polarization also winds a loop along a great circle of the Bloch sphere.
The area subtended by the spin evolution loop in Fig.~\ref{fig:4}(c) is close to a hemisphere [covering the point (0,1,0)] with resultant solid angle of $2\pi\SI[separate-uncertainty=true]{0.992\pm 0.017}{\steradian}$.

We further perform measurements with $\theta_1^\text{in}=\nicefrac{11\pi}6$, $\theta_1^\text{out}=\nicefrac{7\pi}6$, and $\theta_2^\text{in/out}=\nicefrac{\pi}2$. Similar probability distributions after 28 steps of 2DQWs in Fig.~\ref{fig:3} are measured and can be found in Supplemental Material~\cite{supplemental_material}. The spin texture on the outer edge is presented in Fig.~\ref{fig:4}(d). Accordingly, the measured $\braket{\bm\sigma}$
of the spin on the edge is shown in Fig.~\ref{fig:4}(e), with the spin polarization being almost oppositely projected in the $\sigma_x$ basis compared to that in Fig.~\ref{fig:4}(b). As expected, the winding direction of the spin evolution on the Bloch sphere is opposite [Fig.~\ref{fig:4}(f)]. The area enclosed by the spin-evolution loop, with the point (0,1,0) being covered, corresponds to a solid angle of $2\pi\SI[separate-uncertainty=true]{0.999\pm 0.03}{\steradian}$.

Because of EOM signal noise in experiment,
disorder in~$\theta_1$ arises naturally
leading to spins on the edge showing fluctuating projection in $\sigma_x$ and $\sigma_z$ bases~[Figs.~\ref{fig:4}(b) and (e)].
However, protected by chiral-like symmetry, the overall spin winding is robust~[Figs.~\ref{fig:4}(c) and (f)].
To see whether the spin texture depends on the geometry of the edge, inhomogeneous 2DQWs with a $17\times13$ rectangular boundary are implemented.
Similar results that spin polarization changes whenever the walker runs into the corners of the rectangular edge and robust spin winding are observed~\cite{supplemental_material}.

In summary, we have reported the observation of real-space topological spin texture of chiral edge states in a Floquet topological system simulated with photonic 2DQWs. The edge-spin texture in real space also reflects spin-orbit coupling in that the spin changes whenever the walker changes local conducting direction, like turning a corner. Further, we demonstrated control of topology of the edge-spin textures by varying parameters. Our observation shows that, aside from known bulk-edge correspondence connecting bulk topology and the number of edge states,
the edge states exhibit richer exotic features of topological spin texture that correspond to the symmetry of the low-energy Hamiltonian.
We note that, whereas real topological materials could be complicated compared with our Floquet 2DQW model,
our results could inspire new spintronic applications with topology-based devices.

\begin{acknowledgments}
We thank J.Y.\ Zhang, X.L.\ Wang, H.T.\ Wang, W.W.\ Zhang, and W.\ Lin for helpful discussions, and the anonymous referees for insightful comments and very constructive suggestions. This work is supported by the Chinese Academy of Sciences, the Science and Technology Commission of Shanghai Municipality, the National Key R\&D Program of China (Grant No.~2021YFA1400900), National Natural Science Foundation of China (Grants No. 11825401, 11675164,  11921005), the Open Project of Shenzhen Institute of Quantum Science and Engineering (Grant No. SIQSE202003), and the Strategic Priority Research Program of the Chinese Academy of Science (Grant No.~XDB28000000).
\end{acknowledgments}


\end{document}